\documentclass[sigconf,authorversion,nonacm]{acmart}

\usepackage{color}
\usepackage{soul}
\usepackage{url}
\usepackage{fancyhdr}
\AtBeginDocument{%
  \providecommand\BibTeX{{%
    \normalfont B\kern-0.5em{\scshape i\kern-0.25em b}\kern-0.8em\TeX}}}

\begin{document}

\title{Humble Machines: Attending to the Underappreciated Costs of Misplaced Distrust} 

\author{Bran Knowles}
\affiliation{%
  \institution{Lancaster University}
  \city{Lancaster}
  \country{UK}
}
\email{b.h.knowles1@lancaster.ac.uk}

\author{Jason D'Cruz}
\affiliation{%
  \institution{University at Albany, SUNY}
  \city{Albany, New York}
  \country{USA}}
\email{jdcruz@albany.edu}

\author{John T.\,Richards}
\affiliation{%
  \institution{IBM Research -- T. J. Watson Research Center}
  \city{Yorktown Heights, New York}
  \country{USA}}
\email{ajtr@us.ibm.com}

\author{Kush R.\,Varshney}
\affiliation{%
  \institution{IBM Research -- T. J. Watson Research Center}
  \city{Yorktown Heights, New York}
  \country{USA}}
\email{krvarshn@us.ibm.com}

\renewcommand{\shortauthors}{Knowles, et al. \copyright August 2022}

\begin{abstract}
It is curious that AI increasingly outperforms human decision makers, yet much of the public distrusts AI to make decisions affecting their lives. In this paper we explore a novel theory that may explain one reason for this.
We propose that public distrust of AI is a moral consequence of designing systems that prioritize reduction of costs of false positives over less tangible costs of false negatives. We show that such systems---which we characterize as ``distrustful''---are more likely to miscategorize trustworthy individuals, with cascading consequences to both those individuals and the overall human-AI trust relationship. Ultimately, we argue that public distrust of AI stems from well-founded concern about the potential of being miscategorized. We propose that restoring public trust in AI will require that systems are designed to embody a stance of ``humble trust'', whereby the moral costs of the misplaced distrust associated with false negatives is weighted appropriately during development and use. 
\end{abstract}




\maketitle

\section{Introduction}
It is a worrying state of affairs when the majority of the public does not trust AI to make decisions ``about any aspect of their lives'' \cite{BCS} while a growing number of public and private entities are deploying AI to make these very decisions. Understanding and addressing the causes of this distrust is of great importance. While much of the literature conceives of distrust as stemming from a lack of confidence in AI technology to perform correctly, this does not explain the preference for human-based decision making even when it is outperformed by AI \cite{harrison2020empirical}. We propose that something more fundamental is amiss here: the public perceives that AI systems are distrustful \textit{of them}, and they fear they may be unfairly marked as untrustworthy.\footnote{So there can be no confusion, we focus here on distrust of AI by decision subjects. It should not be surprising that this distrust is different to distrust of AI by, for example, decision makers. For a more detailed elaboration of the distinctions between types of trust in AI, see \cite{knowlestrust}.} 

AI classifiers, in the simplest case, determine where an individual falls with respect to a decision boundary. This function may seem value-neutral, even anodyne. But from the vantage of a person who has a vital stake in the decision, the risks of miscategorization are high. Consider: a system deciding who gets a loan is sensitive to indicators of unreliability \cite{codedbias1}; a system deciding who qualifies for welfare looks for evidence of fraud
 \cite{eubanks2018automating}; a system that assesses employee productivity examines evidence of time wasting \cite{verge}; a system designed to ensure public safety looks for indicators of criminality \cite{eubanks2018automating}. When things go right the system assigns the labels \textit{creditworthy, trustworthy, productive, honest} to individuals who merit them. However, individuals with past experiences of miscategorization will reasonably focus on this very real threat.
 
Much of AI's appeal for those who deploy it is that it reduces their overall risk, both the risk of falsely predicting positive outcomes and the risk of falsely predicting negative outcomes. Organizations making decisions in many real-world contexts are focused on the risks of falsely predicting positive outcomes (e.g., extending a loan that ends up in default) because this is costly. However, a system that aims at minimizing the risk of trusting the untrustworthy will expose those who are subject to its decisions to a different kind of hazard, namely the heightened risk of being mistakenly labeled as untrustworthy. Elsewhere this approach has been described as \textit{precautionary}, as opposed to \textit{dauntless}, decision making whereby the costs of false positives are much greater than the costs of false negatives \cite{varshney2016decision}; in that work, it is emphasized that operating in a precautionary regime of decision making presumes a person is guilty (distrusted) until proven innocent (trusted). In this initial exploratory work we consider the \textit{moral consequences of this approach} and examine its contribution to public distrust of AI. 
 
We begin by reflecting on social science literature on the consequences of distrust to understand how distrust\textit{ful} AI may perpetuate distrust \textit{of} AI. Next, we introduce to this domain the concept of ``humble trust'' \cite{d2019humble} and explore its relevance to the goal of promoting public trust in AI. We then consider the technical implications of humble trust and how adopting this development stance may disrupt pernicious distrust in the public-AI relationship. We conclude with an appeal for AI to be designed to balance the immediate and tangible costs of misplaced trust against the intangible but morally significant costs of misplaced distrust.

\smallskip

\textit{Contributions: }
\begin{enumerate}
    \item We develop a novel theory to explain how the default approach for designing AI systems contributes toward public distrust of AI.
    \item We expose the moral costs of this approach, enriching discourses on AI ``harm''.
    \item We offer insights on affirmative measures that might mitigate these harms and improve public trust in AI.
\end{enumerate}
\section{Distrustful AI}

AI systems make decisions that have real world impact on individual people and on groups \cite{Varshney2022}. While these systems are increasingly based on machine learning (ML), other forms of sophisticated processing can underlie these decisions. Put simply, AI ``scores the probability of what you're about to do'' \cite[Cathy O'Neil]{codedbias1}. Predictions about a person's future actions are consequential to deliberations about whether to trust, and we propose that AI engages in a \textit{calculated assessment of one's trustworthiness} with respect to some action that has important consequences to those deploying the system. An individual that is determined to be likely to behave as desired (trustworthy) receives a favorable decision (trusted); whereas one that is insufficiently likely to behave as desired (untrustworthy) receives an unfavorable decision (distrusted).

In social science literature, trust is usually defined as a  \textit{willingness to be vulnerable} to the harm that would occur if an individual acts contrary to expectations (e.g.\,\cite{baier1986trust,lewis1985trust,mcgeer2017empowering,schoorman2007integrative}), and distrust is conceived, then, as ``the retreat from vulnerability''~\cite{d2019humble}. There is an obvious problem in adopting these same definitions of trust and distrust as they relate to AI, namely that it would be strange to speak of AI as being responsive to perceived vulnerability. AI systems do not have mental states like beliefs (about a person's trustworthiness) or affective attitudes (fear, anxiety, etc.). That said, these mental states and attitudes may still be inferred by a human, as there is a strong tendency (by the public as well as the technical community) to anthropomorphize AI \cite{salles2020anthropomorphism}; and they can also quite reasonably be ascribed to the human owners of these AI systems. So when we use the terms ``trust'' and ``distrust'', we are using them, and their relational mechanics, as a useful frame \textit{without} any assertion that AI is capable of thinking or feeling in ways humans do when they trust or distrust. So with that caveat, the vulnerability that an AI would be designed to mitigate is that of the AI owner, who seeks to minimize (typically financial) costs\footnote{We are ignoring the possible benefits to the AI owner, such as the interest earned from a successful loan, in this discussion. But both costs and benefits should be factored in to whether to deploy AI and where to set the decision threshold or thresholds for positive and negative outcomes. In fact, the threshold in a Bayes-optimal likelihood ratio test decision rule is based precisely on the ratio of the cost of false positives and false negatives (along with the ratio of prior probabilities of positives and negatives).} arising from the AI's predictions. We define \textbf{trust}, then, as \textit{a willingness (as defined by some threshold) to accept possible, but unlikely, costs} (i.e.\,if the predicted human behavior proves incorrect). \textbf{Distrust} is \textit{an unwillingness (usually defined by the same threshold) to accept the costs of negative predictions} (i.e.\,if humans behave untrustworthy as expected). Distrustful decisions, therefore, are decisions that mitigate the risks of untrustworthy behavior, in effect avoiding the costs of trusting the seemingly untrustworthy. Seen this way, the outputs of AI systems leading to unfavorable outcomes to individuals (e.g.\,denial of a loan, denial of accommodation, rejection for a job or job interview) can be reasonably interpreted as distrustful decisions, even if they are not statistically unfair.

Inherent to this and all trust relationships is uncertainty. In trusting, ``One is willing to rely or depend on the other although there is always the possibility that he or she will not act as one expects'' \cite{govier1997social}. In other words, trust involves the suspension of (human mental, or in this case, system computational) uncertainty to allow action \textit{as if} the trustee's likely positive or negative behavior is guaranteed \cite{mollering2001nature}. Discussions around AI equity have focused on the issues arising from uncertainty \cite{BuLDPRSW2021}, such as: how AI systems arrive at determinations about the expected behaviors of individuals; what makes two individuals sufficiently similar to assume they might behave the same way if given the same opportunity \cite{MukherjeeYBS2020}; how we might determine whether an outcome for an individual was ``fair'' \cite{HardtPS2016}; and how to go about mitigating biases in decision rules so that predictions are more ``fair'' \cite{WeiNC2021}. Thus far absent is a discussion of the larger consequences of \textit{avoiding the costs of trusting the possibly untrustworthy} to the trust relationship between humans and AI. How does distrustful AI increase the perception of untrustworthy decisions by AI? What is the felt experience of being distrusted, particularly when that distrust feels unwarranted, and/or when experienced in the aggregate across multiple encounters with AI systems? How does a distrustful dynamic perpetuate a cycle of exclusion from opportunity that materially impacts peoples' life prospects \cite{liu2018delayed}? What, then, are the moral stakes involved in this human-AI trust relationship?

\subsection{Distrust contributes to misrecognition of untrustworthiness } 
In \textit{Automating Inequality} \cite{eubanks2018automating}, Virginia Eubanks tells the story of the disastrous implementation of automated  decision-making in the determination of welfare eligibility in Indiana. Motivated to reform what he characterized as a wasteful and fraudulent system, the then-Governor funded the development of a system optimized to reduce costs by identifying individuals who should be moved off benefits. A system predicated on antecedent distrust of the poor---assuming a significant proportion of those requesting welfare are lazy, lying, scroungers---contributed to thousands of people being kicked off welfare despite genuine entitlement because they were mislabeled as uncooperative or undeserving. An automated email informed them of their welfare termination due to ``failure to cooperate in establishing eligibility'' \cite{eubanks2018automating}. For those dependent on this assistance to get by, the decision meant they were unexpectedly unable to afford food and life-saving medical treatment. Such an experience of the dire consequences of being miscategorized as untruthful and undeserving is not soon forgotten. 

This may seem a particularly egregious example of automation gone wrong, but it is far from an isolated incident. For example, more recently the AI-based (self-learning) Dutch childcare benefits system relied on various questionable risk indicators to identify individuals suspected of benefit fraud, leading many to be unjustly penalized, with consequences including massive debt, removal of children from homes, and suicide \cite{Politico}. These examples are instructive about the consequences of AI inclined toward distrust: they are not just more likely to recognize the untrustworthy, they are also more likely to \textit{misrecognize} people as untrustworthy.\footnote{We use the term "misrecognize" as a means of capturing the likely perception of the subject of a negative decision while noting that an AI is more properly described as ``miscategorizing'' the individual.} There are a couple of reasons why this is the case.

\subsubsection{Skepticism and amplification of weak signals} 
Classifying an individual as trustworthy or untrustworthy necessarily entails \textit{construal}: it requires interpretation of signals. For example, how a person dresses can affect whether people are more or less likely to view them as competent; a person's skin color can affect whether a police officer is more likely to view them as a threat. The characteristic stance of individuals in distrustful relationships is \textit{skepticism}, which tends to warp construal in favor of heightened perception of untrustworthiness. We have likely all had the perplexing and frustrating experience of our words being misinterpreted as hurtful by someone who is anticipating us saying something hurtful. When viewed through the lens of skepticism, any data point can be interpreted as evidence of untrustworthiness---e.g.\,an honest mistake, a mere failure to provide evidence in the exact format readable to an AI, or missing a single signature on an application, becomes evidence of a ``failure to cooperate''. Moreover, AI has been found to construe in strange ways, for example interpreting job candidates more favorably in video interviews if they have a bookcase in their background \cite{Newscientist}.\footnote{The new breed of AI tools that purport to infer people's personality traits from signals such as language patterns, vocal tone, and non-verbals (facial expressions and gestures) epitomize algorithmic construal.} 
    
    With construal, there need not be a statistically significant causal     link between the signal and the behavior or quality of interest  (e.g.\,between the presence of a bookcase and a candidate's job capability), and there need not be an explicit encoding of this belief in the algorithm \cite{CuiA2022}; these signals may still factor into the AI's construal of the relevant characteristics of the individual about whom it is making a decision. AI that is optimized to not trust untrustworthy people is \textit{more likely to become hyper-attuned to low-level signals of untrustworthiness} which, in reality, may have little bearing on one's actual trustworthiness. 
    
    Technically speaking, skepticism corresponds to a high cost of false positives relative to the cost of false negatives. This leads to an increased false negative rate (misrecognizing people as untrustworthy; equivalently decreased true positive rate) and a decreased false positive rate (equivalently increased true negative rate). When a machine learning classifier based on correlations rather than causal phenomena operates in the skeptical regime with high costs of false positives, the decision threshold gets pushed to the tail of the likelihood functions where weak correlations reign supreme. Thus, the decision may be based on spurious correlations of the target variable with irrelevant features as they are often weak. This statistical phenomenon and other similar phenomena are explained further in the appendix.

\subsubsection{Decreased responsiveness to evidence of misrecognition} 
    When outcomes do not conform to expectations, this signals the need to revise one’s beliefs. Similarly, when a system produces unexpected outcomes, this offers useful feedback that something may be wrong with that system’s underlying logic, such as a misplaced threshold. But when a system misrecognizes people as untrustworthy, this will only be detected as anomalous if one believes those individuals are proportionally more likely to be trustworthy. In our example above, if those developing or deploying an automated system believe that a vast number of people are scam artists, the fact that a high proportion are found \textit{by the system } to be undeserving of the benefits they are claiming serves to validate the system (albeit using circular logic). Perversely,  \textit{distrust inhibits recognition of misrecognized distrust} by limiting developers' responsiveness to signals of the system itself being untrustworthy. 
    
    Since a Bayes-optimal threshold of a classifier involves the costs of false positives and negatives \emph{and} a prior belief in the proportion of trustworthy and untrustworthy individuals, incorrect prior beliefs lead to greater misrecognition; this effect is amplified when the cost of false positives is very high (the system is precautionary or distrusting) \cite{varshney2016decision}. 

\subsubsection{Combined effect}
Together, these two aspects contribute to \textit{distrustful AI being more likely to be untrustworthy} in its decision making, leading to experiences of direct and sometimes severe harm that end up \textit{justifying public distrust of AI}. Having suffered algorithmic harm through misrecognition by AI, and in many cases never having this injustice remedied, can we blame the public for being wary of other AIs misrecognizing them?

\subsection{People distrust those who distrust them}
\label{sec:peopledistrust}
Having established that distrust makes it more likely that individuals are labeled as untrustworthy, let's now consider what it feels like to be on the receiving end of such labeling. Going beyond the immediate (though certainly real and potentially significant) consequences of the outcome of the decision to the person, it is worth exploring the wider relational dynamics at play. The example of the Post Office scandal in the UK is helpful for exploring these matters as they may pertain to automated decision making. In 2000, in response to concerns about massive financial losses (note the implicit accusation of some mishandling and mandate to find the missing money somewhere), the Post Office implemented accountancy software which began to detect shortfalls at certain branches. The people heading those branches were assumed to have stolen the missing money, and were required to pay the difference---sometimes many thousands of pounds---lost their jobs, and in over 700 cases, were prosecuted for theft, fraud, and false accounting. Though eventually determined to be the result of faulty software, with successful convictions being quashed in 2019, many were irreparably damaged, at least four individuals reportedly taking their own lives~\cite{Eye}.

What we see in this example is that the accusation of untrustworthiness has important effects even, or \textit{particularly,} when it is unwarranted.\footnote{A person who knows they haven't stolen money still does not like being accused of having stolen money. In part this is because, through no fault of their own, they now face the daunting (and sometimes unwinnable) prospect of having to disprove the accusation. Often these battles are not worth fighting, particularly if the distrust is not hugely consequential to a person's life, 
    but this does not mean the injury is forgotten. In the case of the postal workers, many paid the money because it was too difficult to prove they hadn't stolen it; they knew they would not be believed against the word of the automated system.} For our purposes here, whenever one is on the receiving end of an unfavorable decision, that decision carries an accusation of untrustworthiness. Usually\footnote{See our discussion of exceptions to this in section 2.3.} this accusation is bounded by the decision making context \cite{d2018trust}: one might not trust a person with a loan, though this says nothing, necessarily, about how trustworthy they are in other spheres of life (e.g.\,whether they can be trusted to keep their child safe). Nonetheless, it is cause for self-reflection when an algorithm has come to the ``rational'' conclusion that one is not to be trusted. As more and more decisions are being made about us by AIs, it is worth considering what the \textit{cumulative} impact of repeated experiences of being distrusted by AI might be. 

\subsubsection{Resentment} 
If we understand distrust as carrying an implicit accusation, that accusation can be limited to beliefs about one's ability to carry out the desired behavior (though this can certainly be insulting); but when the implication pertains to beliefs about one's benevolence or integrity, this cuts to the moral character of the individual in ways that are deeply injurious. Frustratingly, trustworthiness does not \textit{ipso facto} engender trust. A perfectly legitimate response to one's trustworthiness not being recognized is ill will toward the party who failed to recognize evidence of trustworthiness. And when a person repeatedly experiences a \textit{mismatch between the evidence of trustworthiness required by AI and the evidence they are capable of providing to AI}, this naturally breeds resentment for being made to play a game whose rules are inscrutable and seemingly tilted against them. 
\subsubsection{Demoralization} 
People spend a great deal of time worrying about whether others trust them. This is partly because social strategies in conditions of trust versus distrust differ greatly, and one needs to adapt accordingly; but also, because trust is social currency (it affords us certain opportunities), people are conditioned to value the broadcasting of their trustworthiness to other parties. As such, when a person is trusted, they are motivated to respond to this trust by acting trustworthy so that this signal continues to be broadcast  (see ``trust-responsiveness mechanism'' \cite{pettit1995cunning}). In contrast, a person who is distrusted has less opportunity to broadcast their responsiveness to trust, contributing to what is sometimes called the ``selective labels problem'' in the context of ML that results from unobservables  \cite{ChouldechovaPBFV2018,Wei2021}. Additionally, having less opportunity to broadcast responsiveness to trust is believed to \textit{diminish a person's incentive to act trustworthy} \cite{d2019humble,pettit1995cunning}. It is possible that being repeatedly identified as untrustworthy by AIs may demoralize a person in ways that spill over into how they comport themselves in their daily lives; and in turn, this behavior may produce signals that serve as input to other AIs which are then more likely to identify untrustworthiness.

\subsubsection{Alienation and contempt} 
When a person is distrusted for reasons that appear not to make sense, they may draw on other experiences of pain and injustice in  constructing a plausible explanation. As a result, this can ``open deep wounds that are slow to heal'' \cite{d2019humble}. Similarly, in recognizing a distortion in how one is perceived by AI systems, a person may begin to feel they are different from the machine's ideal type in an unchangeable way that keeps them forever locked out of opportunities that others are getting. In a society where \textit{some individuals are more easily ``readable'' by AI}, who are able to satisfactorily signal their trustworthiness (e.g.\,they produce a sufficiently rich digital trail that matches expected patterns), it's not necessarily the case that a person can change the things about them that make them ``unreadable''. (Note that individuals less easily ``readable'' by AI are those who have had an atypical and/or difficult life; and being ``unreadable'' by AI may compound the struggles they already face.\footnote{This is of course relevant to concerns regarding AI's comparatively poor performance (e.g.\ face recognition, pedestrian detection, and disease diagnosis) for marginalized groups.}) Those experiencing exclusion because they cannot satisfy AI systems may quite rightly feel contempt for the entire world represented by ``AI''---the system of systems that circumscribe the new rules in society that appear to make it harder for certain individuals to succeed. 

\subsubsection{Combined effect}
These mechanisms help explain how \textit{distrustful AI provokes reciprocated distrust by the public}. There is an important affective dimension of trust and distrust---while one may reason about trust, emotions have a strong influence on that reasoning \cite{jones2019trust}. For example, ``Resentment leads us to focus and sometimes obsessively replay the wrong that has been done and cuts off search for possible mitigating factors or alternative explanations of it'' \cite{jones2019trust}. Contempt\footnote{Others have noted a difference between ``passive contempt'', resulting in an individual's disinterest in the object, and ``active contempt'', resulting in heightened attention on the object as a threat (see \cite{bell2013hard}). Here were are referring to active contempt by the public.} is a ``totalizing emotion'': ``it focuses on the person as a whole rather than on some aspect of them'' \cite{jones2019trust}. This is important when we consider people's distrust of AI to make \textit{any} decision about them \cite{BCS}, as this reflects a collapsing of all AIs into a single threatening entity. Contempt feeds further distrust (it is ``distrust-philic'' \cite{jones2019trust}) because it ``pre-empts finding grounding for trust in even the smallest of domains,'' and ``undercuts optimism about both the competence of the other and their willingness to be responsive to our dependency'' \cite{jones2019trust}. For the purposes of our discussion, it is important that these emotions reduce the individual's receptivity to the notion that AI \textit{could} be trustworthy, e.g.\,even if improvements are implemented which correct for the original misrecognition. (We note that this has important implications regarding the limits of explainable AI as a corrective intervention to promote public trust.)  Misplaced distrust by AI not only justifies rational, cognitive distrust of AI by the public (section 2.1), it also \textit{triggers affective feedback loops that intensify and entrench public distrust}. 

\subsection{The inertia of distrust}
So far we have noted how a perceived distrust \textit{by} AI can breed public distrust \textit{of} AI. We turn now to exploring feedback mechanisms that can fuel deeper or more widespread distrust on the part of AI, perpetuating an ever worsening public-AI relationship. Credit scoring offers a tangible example of this. Ideally, trust/distrust is bound by a context (section 2.2), but credit scores are an indicator used for many purposes beyond strict determination of credit worthiness. A low credit score can make it harder for people to get a mobile phone, or rent an apartment. It can lead to higher car insurance premiums. Employers may run a credit check on candidates, even for jobs that do not require the direct handling of money. It can block security clearances that affect military service members \cite{ConsolidatedCredit,Investopedia} or other government workers. Bad credit can even reduce a person's romantic prospects---it may signal a potential risk (e.g.\,becoming financially entangled with someone whose bad credit may spread), or be seen as a ``red flag'' hinting at other issues, deterring continuance of early-stage romances \cite{Forbes1}. Some of these impacts can, in turn, feed back to further lower a credit score. Being turned down for jobs is no help whatsoever to a person trying to establish good credit, and not having a romantic partner can reduce stability to buffer unexpected financial turbulence. Moreover, credit scores are frustratingly unstable---a person may have a lifetime of financial reliability, but if they abruptly begin to miss payments because of a setback beyond their control, say they are hospitalized with an illness or suffer some other unanticipated disruption in their life and are unable to make payments on time, their score may drop appreciably.

It is a known phenomenon of interpersonal relationships that distrust tends to be self-reinforcing \cite{mcgeer2002developing} (explored at length in \cite{d2019humble}), but this is understood to be at least partially fueled by distrust-philic emotional responses in both directions. Interestingly, here we see that AIs, entirely lacking in affect, are also implicated in what can be seen as distorted reasoning that \textit{reinforces untrustworthiness classifications}. There are three components of this distortion.

\subsubsection{Fundamental attribution error} 
     The reason credit score is such a widely used indicator is because it offers a seemingly objective measure of a person's moral character. In reality, there are any number of mitigating circumstances for a person's low credit score, but when an AI uses credit score as a feature in areas outside of credit worthiness, it is assuming that a) the score conveys relatively stable information about the person's disposition, and b) that this characterization is useful in predicting behavior across a wide range of contexts. In social psychology, when people over-emphasize disposition (over the influence of situation) in understanding a person's behavior it is known as the \textit{fundamental attribution error} \cite{jones1967attribution}. It would, of course, be impossible for AIs to process all of the potentially relevant situational contributors to a person's behavior. Our concern is not with this particular limitation, as it is the very limitation that necessitates trust in the first place (i.e.\,the fact that one can never be certain of a person's future action). Instead, we note that when an output of one system is used as a feature for another, whatever situational information that may have influenced the first system is at least diluted if not eliminated in the second system. This leads to a systematic over-emphasis on disposition which ultimately means that 
     \textit{a person who is misrecognized as untrustworthy by one AI has a greater chance of being misrecognized as untrustworthy by other AIs}.
\subsubsection{Reliance on proxies} 
    As we defined above, distrust by AI manifests in an unfavorable decision to an individual. A person who is trusted, in contrast, not only receives a favorable decision, but that decision often creates opportunities that are characteristic of ``trustworthy'' individuals. For example, if a person's credit score is sufficiently high that they are granted a mortgage to purchase a house, they may now be categorized as a ``homeowner''; their status as homeowner might be interpreted as indicative of associated trustworthy characteristics (e.g.\,``responsible'').\footnote{Though clearly problematic, this kind of inference is common with AIs. For example, Alibaba uses payment history to piece together a story about an individual's moral character: ``Someone who plays video games for ten hours a day, for example, would be considered an idle person, and someone who frequently buys diapers would be considered as probably a parent, who on balance is more likely to have a sense of responsibility'' (Li Yingyun, Sesame’s technology director; in \cite{NewYorker2}).} Consider the lasting impacts of a decision regarding eligibility for rented accommodation: A trusted person is granted an apartment in an affluent neighborhood. A distrusted one is denied the same accommodation, so takes an apartment in a less affluent neighborhood, which (not unrelatedly) has a higher crime rate. In a future decision about these two individuals, the one with a ``better'' postcode is more likely to be seen as more trustworthy \cite{pope2011implementing} to the extent that proxy is used as a model feature in other AIs.     To take another example, a person who is unemployed for more than six months can be categorized as ``long-term unemployed'' (see \cite{marks2022algorithmic}), the implication being that there is some underlying flaw in motivation or work habits that justifies their exclusion from consideration for a job. Resume filtering AIs that consider such a categorical feature would embody the assumption that if others have not employed this candidate, there must be a reason not to employ them.     The overall point here is that decisions based on proxies that are seen as being evidence of untrustworthiness---living in a ``bad'' postcode or having a gap in employment---perpetuate long-term disadvantage by \textit{making it both easier for the trusted to be recognized as trustworthy and harder for the distrusted to be recognized as such.} 
    
\subsubsection{Asymmetrical feedback on responsiveness to trust} 
    When a person is trusted, they are given an opportunity to carry out the task they are trusted to do. This generates new data that serves as ``confirming'' feedback to the trustor as they perform ongoing re-calibration of trust \cite{Wei2021}. An individual's tendency to meet commitments or to fall short of them will result in adjustments to an AI's classification of their trustworthiness. It is not the case, however, that the untrusted have equal opportunity to become reclassified. A consequence of being distrusted is that one cannot then demonstrate how they would have responded to trust. Again, people are motivated to respond to trust by being trustworthy \cite{pettit1995cunning}. They may also draw strength from people's hopeful vision of them as trustworthy, encouraging them to be more trustworthy than they otherwise might be \cite{mcgeer2008trust}.\footnote{See also the ``self-concept mechanism'' \cite{alfano2016friendship}, whereby another's trust enables a person to see themselves as trustworthy and is thus motivated to act in ways that allow them to retain this positive self-concept.} Distrustful strategies---retreating, withdrawing, avoiding reliance---lead to systematic under-trusting \cite{fetchenhauer2010so} by \textit{reducing information about people's trust-responsiveness} that is needed to recalibrate misplaced distrust.

\subsubsection{Combined effect}
As in interpersonal relationships, distrust can create \textit{pernicious spirals} that are very difficult to escape. Distrust by AI feeds itself insofar as it leads to a system (by which we mean both within-system and system-of-system) feedbacks that \textit{prevent re-trusting those classified as untrustworthy.} This is partly what is so concerning about China's Social Credit System, which attempts to promote societal trust by ``allow[ing] the trustworthy to roam everywhere under heaven while making it hard for the discredited to take a single step'' \cite{mistreanu2018life}. A small misstep, being discredited in one small way, can have cascading consequences that feed a downward spiral---a phenomenon caricatured in the \textit{Black Mirror} episode entitled ``Nosedive''.\footnote{The episode also illustrates the tendency for these scoring systems to promote a gaming stance by those being scored---the protagonist in this episode is singularly focused on doing things that will increase her score so that she can get into a high-end, exclusive apartment complex. Even if one were to purposely produce data designed to signal their trustworthiness to AIs (e.g.\,paying an advisor or agency to ``boost'' credit scores) in order to escape from a distrust spiral, a gaming strategy has potential to backfire. For example, if algorithms are modified they may incorporate new indicators of trustworthiness that a person focused on a gaming strategy will not be actively cultivating, or may even interpret evidence of gaming as evidence of untrustworthiness. At a minimum, we can say this cat-and-mouse game is not a route to mutual trust between people and AI.} Our concern is less about the potential of catastrophic losses of trust in a given individual (this is still somewhat unrealistic, even in China), and more that the difficulty an individual faces in re-establishing their trustworthiness makes misrecognition by AI more consequential than it may immediately seem.

\section{A proposal to emulate Humble Trust}

The dysfunction we have described above can be summarized as \textit{public anxiety that they will be misrecognized by AI as being untrustworthy}. As we have shown, the public's distrust of AI is warranted; they may not be trusted by AI even when they are trustworthy, and the consequences of this can be significant. So far we have explored some of the reasons why this may occur and how they relate to our characterization of AI as ``distrustful''. In what follows we outline some affirmative measures that might disrupt this dynamic and foster greater public trust of AI.

We take our inspiration from a stance called ``humble trust'' \cite{d2019humble}. Underlying humble trust is a dedication to the moral obligation by those in positions of power to care about the harm caused by misrecognition. This means balancing the aim to not trust the untrustworthy with the aim to \textit{avoid misrecognition of the trustworthy}.
The practice of humble trust entails \cite{d2019humble} (emphases added): 
\begin{enumerate}
    \item ``\textbf{skepticism} about the warrant of one's own felt attitudes of trust and distrust'';
    \item ``\textbf{curiosity} about who might be unexpectedly responsive to trust and in which contexts''; and
    \item ``\textbf{commitment} to abjure and to avoid distrust of the trustworthy''.
\end{enumerate}

In what follows we explore each of these in turn and offer an initial sketch of what it would mean to design more ``humble'’ machines. We relate the three aspects to (1) features, (2) labels, and (3) costs and thresholds of the decision functions in a machine learning system.

\subsection{Skepticism: Confidence and verification}

In situations where an outcome is certain, trust is irrelevant \cite{lewis1985trust}. It is nonsensical, for example, to speak of ``trusting'' that an object will fall; one \textit{knows} it will fall; gravity (at least in the absence of unusual countervailing forces) guarantees this. But trust is key in interpersonal relationships because a person’s future behavior is rarely knowable with certainty. The best one can do is make an informed prediction based on evidence about past behavior by the same person and other similar people. It is sensible to want to discern with the greatest possible accuracy who is trustworthy and who is not, but in doing so, one needs to be open to the possibility of getting it wrong. While machines are capable of high-speed processing of large volumes of data (in this case, evidence relating to trustworthiness), we know that machines' interpretation, like humans’, is highly susceptible to bias. It is not always clear which feature or combination of features is most predictive of the desired behavior, nor how the available data relates to those features. Key to avoiding overestimating the predictive capability of machines is   \textit{recognizing the information loss that occurs in selecting a set of features while ignoring others}.\footnote{The loss of information is a must due to the data processing inequality of information theory.}  What is the system not seeing by focusing on what it is focusing on? An appropriately skeptical stance would be to assume the model is missing information that could be relevant.

    Returning to the example of credit scoring, one approach to being receptive to additional information is (perhaps unsurprisingly) allowing an individual to provide additional information. Credit rating agencies allow a Notice of Correction to be added to a credit report. It is limited, however, to 200 words. Ensuring it has been added to all credit agencies is tedious. Moreover, it may, in some cases, lead to more scrutiny of the adverse markers being noted in the correction. To our knowledge, there is also no automatic processing of such notices. So it is simply part of the credit report that can be viewed by entities assessing credit worthiness. Whether this is likely to heal a dysfunctional trust dynamic hinges on a Notice \textit{being heard} and the evidence of that hearing being available to the individual who submitted the Notice. It also requires individuals to be both knowledgeable and proactive regarding their credit score, which is typically more challenging for exactly the same people who would be more likely to be automatically excluded based on bad credit. But at least this opens the door to a conversation (in theory), and it represents an explicit acknowledgment that we cannot be satisfied that the score alone tells the whole story.

More generally, the absence of relevant features leads to uncertainty in the data and machine learning model. It also prevents valid causal modeling because the \emph{ignorability} assumption of causal inference (all confounding variables available as features) is violated. Including more informative features results in less uncertainty and better discrimination between trustworthy and untrustworthy people. As illustrated in the appendix, this manifests as narrower likelihood functions that are better separated from each other. Skepticism implies aiming for ``better discrimination of untrustworthiness itself from the illusory appearance of untrustworthiness'' \cite{d2019humble}.

Moreover, as discussed in section \ref{sec:peopledistrust}, people who are less ``readable'' have atypical ways of demonstrating their trustworthiness. The onus is then on the skeptical AI system and its developers to actively seek out non-traditional evidence of trustworthiness as features that allow people to show themselves. For example, AI-based credit decisions for unbanked customers in Kenya and other parts of East Africa use mobile phone-based features \cite{SpeakmanSM2018}. Also consider the approach adopted by Upstart.com, which looks to extend credit to the traditionally non-creditworthy by explicitly looking for other signals of trustworthiness. Not only does this give people a way in to earning trust, it taps into a market segment that would be particularly responsive to trust---they are highly motivated to demonstrate their trustworthiness to establish better credit. 

The ``Rooney Rule'' in hiring professional football coaches that requires teams to interview at least one minority candidate provides additional informative features that would not be obtained if the interview were not conducted. There is an awareness that AIs need to be ``fair''. We would argue that fairness, in essence, means that if a person is trustworthy they are recognized as such---in this case, someone who is capable of doing the job they are applying for would get an interview. An active feature acquisition approach proposed by Bakker et al.\ operationalizes skepticism exactly along these lines and achieves fairness for both groups and individuals by continuing to seek additional features about individuals as long as the AI remains too uncertain \cite{BakkerTGPVW2021}.\footnote{Group fairness, such as the concept of equality of opportunity requires average true positive rates to be equal for groups delineated by protected attributes such as race and gender. Individual fairness requires similar individuals to receive similar predictions.}

\subsection{Curiosity: Trust-responsiveness}

Part of being open to the possibility of having gotten a decision wrong is being curious about what might have happened if a different decision were made: the counterfactual situation, if you will. This means creating opportunities for the AI to learn about the trust-responsiveness of people who fall below the decision threshold---in practice, extending trust to those who might betray it and seeing if the expectation of betrayal is fulfilled.\footnote{In the context of designing systems to support cooperation between parties, this approach to trust-building has been described as giving people ``chances to fail'' \cite{knowles2015models}.} As mentioned in section \ref{sec:peopledistrust}, not doing so is known as the ``selective labels problem'' \cite{Wei2021}. Indeed, given that a person's trustworthiness is partially a function of social affirmation \cite{d2019humble} (see section \ref{sec:peopledistrust}), the salient question in determining who to trust is not \textit{is this person trustworthy?} but rather \textit{is this person trustworthy when trusted}? 

For example, curiosity about trust-responsiveness in algorithmic hiring might entail taking some portion of rejected candidates back into the pool and feeding data regarding their subsequent performance at interview (or if ultimately offered the job, their job performance) into the model to refine the AI's rejection criteria. The way to imagine such a process (illustrated in the appendix) is by expanding a decision threshold into a band---a sort of gray area---where the AI is most uncertain. Within this band, the trustworthy/untrustworthy determination is randomized. 

More nuanced solutions to the selective labels problem have been addressed in principled ways in the decision making and machine learning literatures. For example, Wei balances the costs of learning with future benefit through a partially-observed Markov decision process that shifts a classifier's decision threshold to more and more stringent positions as it sees more people that would normally have been classified as untrustworthy \cite{Wei2021}. This may be seen as a way of conducting \emph{safe exploration} wherein the system exhibits curiosity up to a point that does not induce undue harms \cite{MoldovanA2012}. It may also be seen as involving \emph{satisficing} behavior: a decision strategy that aims for a satisfactory result, but not necessarily the utmost optimal one if curiosity were not a consideration \cite{Wierzbicki1982}.

\subsection{Commitment: Investing in identifying trustworthiness} 

One way of understanding distrustful AI is the relative prioritization of cost reduction (again ignoring benefit maximization) of false positives over false negatives. Where the decision threshold is placed has a direct influence on the balance between these two kinds of errors. But regardless of where the threshold is placed, the overriding objective in deploying an AI in most instances is \textit{lowering the costs of making any decision}. Eliminating humans from the decision process is tempting for this very reason. But AIs can do more harm then good when this sort of efficiency is pursued to the exclusion of other values, such as quality and fairness. Take the example of automated resume filtering: the temptation is to design the AI to ``winnow down the number of applicants to the most qualified few, in the shortest possible time, and at the least cost\ldots [but this] will necessarily exclude some very strong candidates, who may only need a little extra training to excel in the roles the companies are seeking to fill'' \cite{marks2022algorithmic}. Commitment is exemplified by doing something \textit{even when it is tempting not to}; so a commitment to avoiding distrust of the trustworthy means making adjustments to the model or wider decision making process even when those changes reduce the overall efficiency. 

Before an AI is built, a deliberative process can weigh the broader implications of its deployment. AI Ethics Review Boards are beginning to appear in multiple organizations with the remit to consider possible harms to individuals (both clients who might be adversely impacted and employees whose jobs might be altered or displaced) and to the organization's external reputation. Since a high cost of false positives relative to the cost of false negatives is implicated in the misrecognition of untrustworthiness, more evenly balancing the two costs is a clear path toward commitment. This serves to bring the decision threshold to a more trusting position. 

Related, and also related to the randomization method to achieving the curiosity stance of humble trust, is making a commitment to ``selective classification'', also known as ``classification with a reject option'' \cite{BartlettW2008}. In this paradigm, when the AI lacks confidence and is unsure whether a person is trustworthy or untrustworthy, it passes the decision on to a human decision maker (whose time and effort is costly). In practice, this amounts to the AI not classifying people who fall in a band around the decision threshold (this is the same kind of band used in randomizing the decision). Human decisions can also be fed back into model improvements, adjusting thresholds or providing hints of additional features of merit. Moreover, the existence of this human oversight can be made visible to those subject to AI decisions, in some cases involving a dialog between the human evaluator and the decision subject. 

For paradigms such as selective classification to be tenable, it is critical that first, the AI system provide an indication of its confidence (this is known as uncertainty quantification \cite{bhatt2021uncertainty}) and that second, the confidence is well-calibrated so that it is not over-confident or under-confident \cite{BuLDPRSW2021}. (Modern neural networks are notoriously over-confident \cite{GuoPSW2017}.) Quantifying uncertainty is in itself a commitment to humility as is the provision of understandable explanations of a decision. Explanations of a negative decision, even explanations generated by the AI itself, can be cast as suggestive rather than definitive, and would ideally provide information about how the decision can be appealed or changed down the road through attention to one or more of the features that most contributed to the outcome. Finally, after deployment, ongoing monitoring can evaluate whether and why individuals are receiving negative decisions. This can expose areas of potential weakness in the model, supporting a continual process of improvement.

\section{Discussion}
Recent years have seen a growing interest in understanding and preventing negative consequences of AI. So far, considerations of harm have been focused on adverse impacts on an individual or group \textit{at-the-point-of-decision} (generally assessed by one or more computed metrics comparing members of so-called protected groups with unprotected ones). 
In this paper, we have shown that each instance of misplaced distrust (harm) by AI generates momentum in the direction of further harm to the individual, and by extension, the group to which the individual belongs. An automated decision is not merely an outcome; crucially, it is a \textit{signal}---both to the decision's subject and to other decision makers (including AIs)---that can reduce the subject's likeliness of being perceived favorably in the future.

Many of the mechanisms we have described in this paper presuppose that the subject is aware that they have received a negative decision by an AI. This is often the case, but harm can occur even when the subject is not aware. While laws exist to protect individuals against discrimination, for example, ``knowing where to look and obtaining relevant evidence that could reveal \textit{prima facie} discrimination will be difficult when automated discrimination is not directly experienced or `felt' by potential claimants'' \cite{wachter2021fairness}. At the same time, we are struck by the fact that \textit{sometimes} being miscategorized reveals (in a rather unflattering light) the AI underlying a decision, and in doing so, causes people to actively examine their trust in AI. Pioneering trust theorist Annette Baier writes, ``We inhabit a climate of trust as we inhabit an atmosphere and notice it as we notice air, only when it becomes scarce or polluted'' \cite{baier1986trust}. Surely in such instances of negative exposure, ``failure to trust the trustworthy'' \cite{jones2013distrusting} is significantly more costly than trusting some portion of untrustworthy people, as it can lead to irreparable loss of trust in the organization deploying the AI, and may contribute to more diffuse lack of trust in any future use of AI.

While there are, then, multiple reasons to be concerned with the costs of misplaced distrust---and we encourage organizations to be cognizant of these when balancing their costs---it is difficult to link the financial consequences of misplaced distrust directly to an organization's ``bottom line''. The case is clearer from a moral perspective; prioritizing cost minimization within the current paradigm leads to miscategorization, misrecognition, and perpetuation of marginalization and distrust. Careful adjustment of AI decision thresholds, and openness to a broader range of features signaling an individual's trustworthiness, can mitigate some of these effects, but it will take more than this to restore the public's trust in AI. The public will need to see that those who develop and deploy AI systems genuinely value the people who are affected by AI decisions, that they exhibit empathy, overriding a mere desire for efficiency, and that they are committed to earning the public's trust. Taking the long view, we cannot have a thriving society if the public distrusts a key component of our technical infrastructure.

\section{Conclusion}
Humble trust clearly does not imply trusting everyone. Rather, it calls for developers and owners of AI systems to be aware of the fact that they do not know all relevant characteristics of each individual subject to an AI decision. It further encourages them to look for (and provide opportunities for the future generation of) new signals of trustworthiness, thereby improving their ability to recognize the trustworthy. Finally, it suggests they look beyond the immediate efficiencies of decision making to consider the long term harms (both to individuals and AI-deploying institutions) of careless classifications of untrustworthiness.

While many applications of AI pose risks of exacerbating unjust distributions of trust, and therefore of opportunity, AI also offers unique mechanisms for disrupting pernicious equilibria of distrust. It is possible to calibrate decisions made by AI systems with tools that are unavailable to the calibration of our own psychologies. Human attitudes of trust and distrust can be altered indirectly, but they are not under a person's direct voluntary control. As such, it is difficult for us to adjust our personal attitudes of trust and distrust to align with our moral aims. The developers of AI systems do not face this problem. With work, AI can be brought into better alignment with these aims.

\appendix

\section{Statistical Illustrations}
In this appendix, we will illustrate how the AI's distrust of people manifests in a statistical sense and how humble trust solutions may be interpreted. Consider a classification problem with a vector of features $X$ and a binary label $Y \in \{0,1\}$. The value $Y=0$ is a \emph{negative} and the value $y=1$ is a \emph{positive}. The task is to find a classifier $\hat{y}(X)$. The main tools for the illustration are the likelihood functions $p_{X\mid Y}(x\mid y = 0)$ and $p_{X\mid Y}(x\mid y = 1)$, prior probabilities $p_0 = p_Y(y=0)$ and $p_1 = p_Y(y=1)$, and costs $c(Y,\hat{y}(X))$, where the cost of a false positive is $c(0,1) = c_{01}$ and the cost of a false negative is $c(1,0) = c_{10}$. The Bayes-optimal classifier is:
\begin{equation}
    \hat{y}(x) = \begin{cases} 0, & \frac{p_{X\mid Y}(x \mid y = 1)}{p_{X\mid Y}(x \mid y = 0)} \le \frac{c_{10}p_0}{c_{01}p_1}\\
        1, & \frac{p_{X\mid Y}(x \mid y = 1)}{p_{X\mid Y}(x \mid y = 0)} > \frac{c_{10}p_0}{c_{01}p_1}
    \end{cases}.
\end{equation}

In the figures, the blue curve is an example of a likelihood function $p_{X\mid Y}(x\mid y = 0)$ (people who are truly untrustworthy) and the orange curve is an example of a likelihood function $p_{X\mid Y}(x\mid y = 1)$ (people who are truly trustworthy). The area of the blue-shaded region is the false positive rate and the area of the orange-shaded region is the false negative rate. The example shows a one-dimensional $X$, but the intuition holds for higher-dimensional feature vectors; in fact, we can imagine this $X$ to be a sufficient one-dimensional projection of all the features.

\begin{figure}
    \includegraphics[width=0.9\columnwidth]{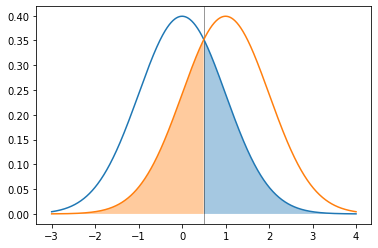}
    \caption{Equal costs for false positives and false negatives.}
    \label{fig1}
\end{figure}
\begin{figure}[h]
    \includegraphics[width=0.9\columnwidth]{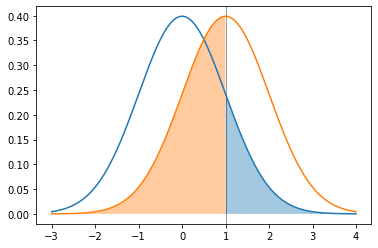}
    \caption{Greater cost for false positives than false negatives yielding a larger false negative rate and smaller false positive rate.}
    \label{fig2}
\end{figure}

In Figure \ref{fig1}, when $c_{01}$ and $c_{10}$ are equal, the decision threshold is in the middle, and the false positive rate and false negative rate are equal. In Figure \ref{fig2}, the AI is more distrustful and thus has $c_{01} > c_{10}$. 
\begin{figure}
    \includegraphics[width=0.9\columnwidth]{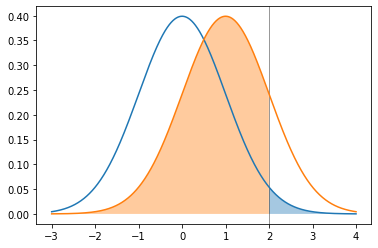}
    \caption{Even greater cost for false positives than false negatives yielding an even larger false negative rate and even smaller false positive rate.}
    \label{fig3}
\end{figure}
This pushes the decision threshold to toward the right and yields many more people being classified as untrustworthy. Since the area of the orange region is now more than the area of the blue region, more people are also being misclassified as untrustworthy. In Figure \ref{fig3}, the AI is even more distrustful and thus has $c_{01} \gg c_{10}$ and the situation is even more exaggerated toward many people being misclassified as untrustworthy. Also notice how the decision threshold is way to the right and at a point in the tail of the likelihood functions, which corresponds to weak and less informative data.

\begin{figure}
    \includegraphics[width=0.9\columnwidth]{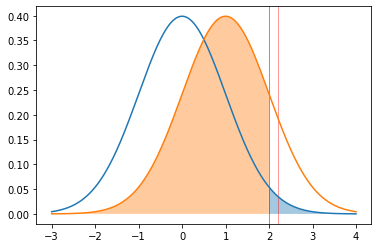}
    \caption{The red line indicates a decision threshold corresponding to the AI having mistaken prior probabilities for the classes.}
    \label{fig4}
\end{figure}

Figure \ref{fig4} illustrates a situation in which the AI has mistaken prior beliefs $p_0$ and $p_1$ on the prevalence of untrustworthy and trustworthy people in the population. The red line is a threshold resulting from the mistake and causes even more misclassification than would happen otherwise. When $c_{01} \gg c_{10}$, that increase in misclassification is relatively large and also pushes toward more false negatives.

\begin{figure}
    \includegraphics[width=0.9\columnwidth]{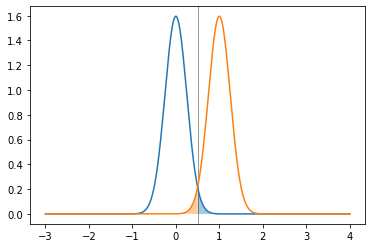}
    \caption{Lowered uncertainty through more informative features.}
    \label{fig5}
\end{figure}

In pursuing humble machine trust, seeking out and including more informative features reduced uncertainty and better discriminates the untrustworthy from the trustworthy. This is shown as narrower likelihood functions in Figure \ref{fig5} in which both the false negative rate and false positive rate (areas of the orange and blue regions) are smaller compared to Figure \ref{fig1}.

\begin{figure}
    \includegraphics[width=0.9\columnwidth]{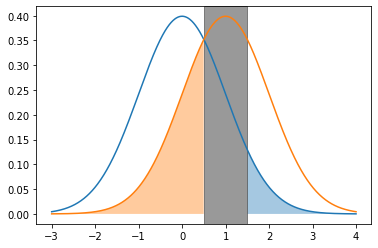}
    \caption{The gray band around a decision threshold may be used for randomization or to revert to a human decision maker.}
    \label{fig6}
\end{figure}

\begin{figure}
    \includegraphics[width=0.9\columnwidth]{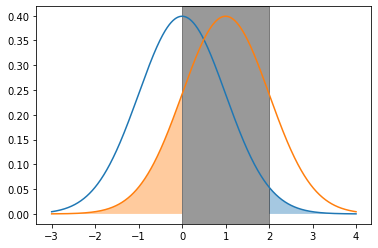}
    \caption{The randomization or reject option region is wider.}
    \label{fig7}
\end{figure}

\begin{figure}
    \includegraphics[width=0.9\columnwidth]{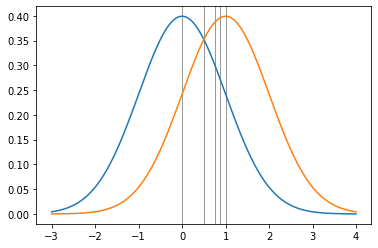}
    \caption{The decision threshold progressively moves from left to right, starting in a humble position.}
    \label{fig8}
\end{figure}

Toward curiosity, Figure \ref{fig6} shows a gray band around the decision threshold in which the decision of trustworthy and untrustworthy may be randomized. An even wider gray band is shown in Figure \ref{fig7} that yields even more randomization. Figure \ref{fig8} illustrates how the curiosity-driven solution of \cite{Wei2021} progressively moves the decision threshold from left to right and is more humble at the beginning. The gray bands in Figure \ref{fig6} and Figure \ref{fig7} may also be used as places where the decision $\hat{y}(X)$ reverts to a human decision maker. Notice that in Figure \ref{fig7}, the AI's false positive rate is the same as in Figure \ref{fig3} (area of blue shading) with much smaller false negative rate (area of orange shading).

\section*{Acknowledgments}
This work is partially funded by the SUNY-IBM AI Research Alliance under grant number AI2102, the ESRC funded grant BIAS: Responsible AI for Labour Market Equality (ES/T012382/1), and by the Data Science Institute at Lancaster University.

\balance{}
\bibliographystyle{ACM-Reference-Format}
\bibliography{sample-base}

\end{document}